# Modeling of Mechanical Overlap Joints in Magnetic Shields

Anthony C. Crawford    Fermilab Technical Div. / SRF Development Dept.    acc52@fnal.gov    09Oct15

This study determines a useful value to use for the gap width of mechanical overlap joints in models of magnetic shielding. The average value of 0.1 mm is found to agree with measurements.

The accuracy of calculated values produced by analytical magnetostatic models is dependent on the accuracy of input parameters. The purpose of the present work is to evaluate the effectiveness of typical mechanical joining techniques for high permeability magnetic shields and to determine a realistic value for the width of the mechanical gap at overlapped joints. This value would then be used as a parameter for modeling shield configurations for future applications. Here we treat the case for cylindrical shields made up of two half shells. The use of split cylinders is frequently required by assembly on to a cylindrical object. The particular overlap joint under study is typical for cylindrical shells used to shield superconducting RF cavities.

**The Model**

The two dimensional magnetostatic program FEMM (Finite Element Method Magnetics) [1] was used for calculations. The cylinder in the model is infinitely long. Input parameters are shown in Table 1.

| Parameter | Value | Units |
|---|---|---|
| Shield Relative Permeability | Constant for each calculation | dimensionless |
| Shield Thickness | 1.0 | millimeter |
| Shield Radius | 125 | millimeter |
| Transverse Ambient Field | 437 | milliGauss (100% Vertical) |
| Gap Size | 0.05 to 0.5 | millimeter |
| Overlap Length | 25 | millimeter |

Table 1.   Input Parameters for the Model

We first consider the worst case scenerio for leakage field at a mechanical joint: a cylinder that is split in the horizontal plane that is exposed to a vertical ambient magnetic field. The basic geometry and mesh configuration is shown in Figures 1 and 2. Magnetic field lines and a color contour plot of the calculated field are shown in Figures 3 and 4.

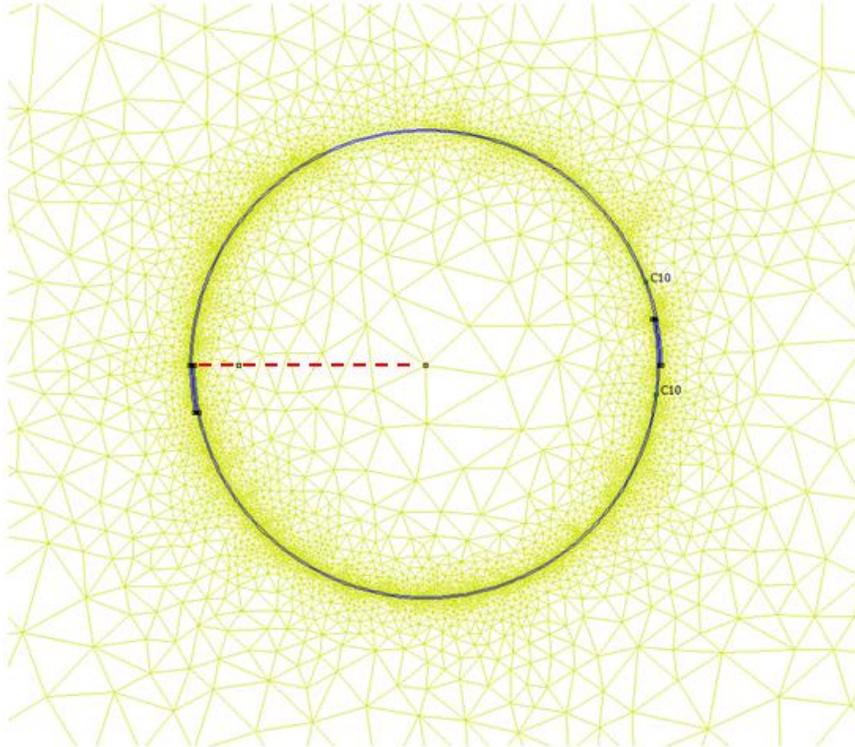

Figure 1.    The Geometry and the Mesh

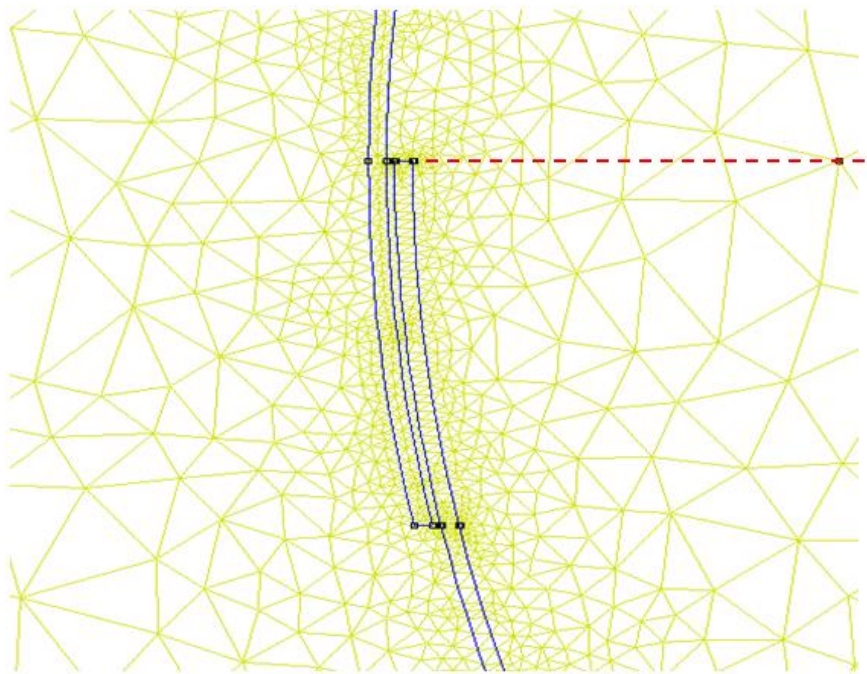

Figure 2.    The Mesh at the Overlap Joint
( 0.5 mm Gap Size X 25 mm length shown)

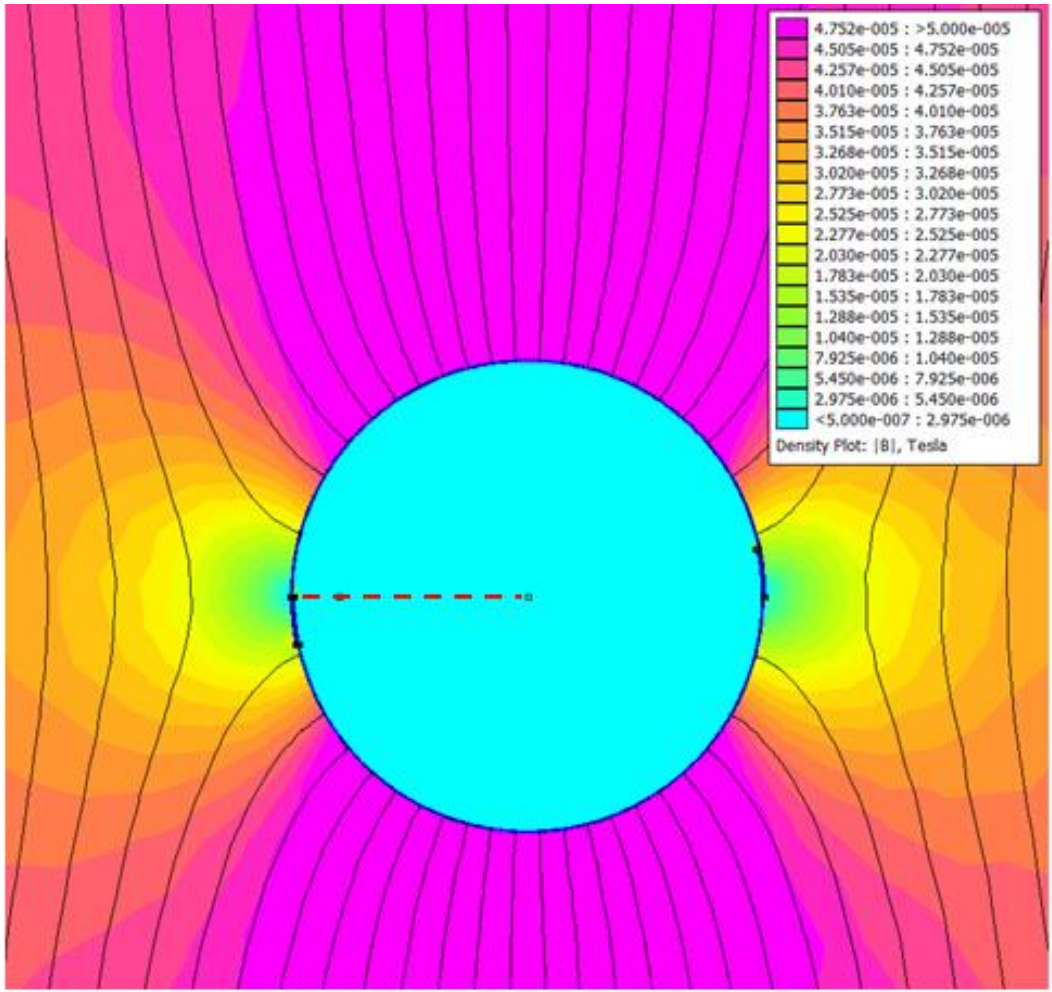

Figure 3. The Case of Immersion in a 437 milliGauss Vertical Ambient Field
(Calculation for 0.1 mm Gap, $\mu_r$ = 24000 shown)

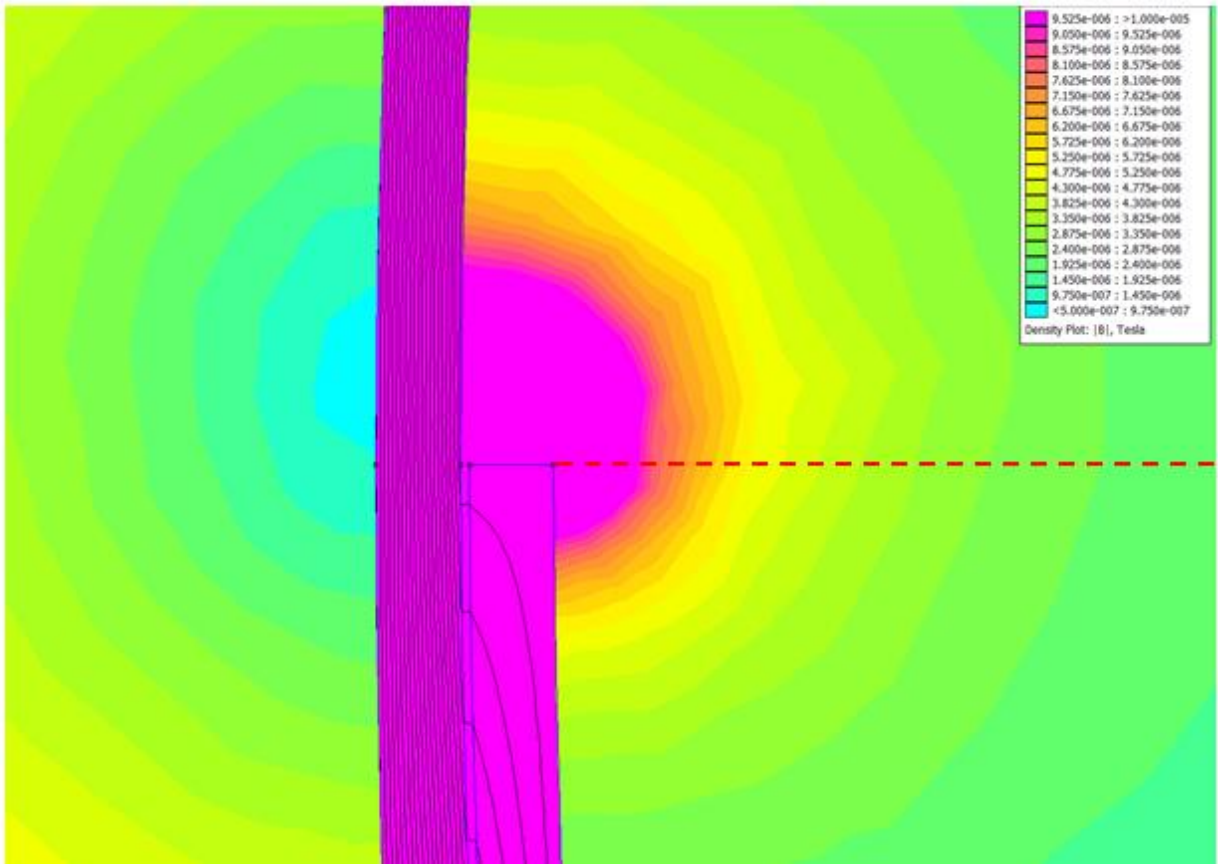

Figure 4.    Enlarged View of the Joint Area From Figure 3

**Measurements**

A Fermilab International Linear Collider (ILC) cryomodule magnetic shield assembly was used for measurements. Overlap joints are compressed by means of screws placed at the centerline of the joint at 125 mm intervals along the axial direction of the cylindrical geometry, as shown in Figure 5. The result is a joint that has good contact at the screw locations, but with a varying gap width at locations between two screws.

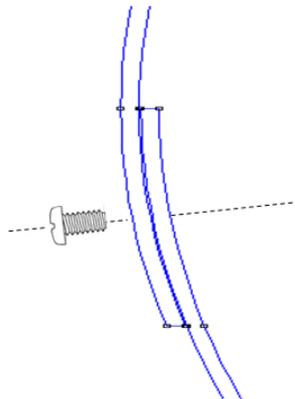

Figure 5.   The Overlap Joint is Compressed by Screws

The axis of the cylindrical shield was in the horizontal plane. The cylinder was located so that the transverse horizontal field component was close to zero. Measurements were made along the radially oriented dotted red line shown in Figures 1 through 4. All measurements were made at room temperature and at a distance inside the end of the cylinder approximately equal to the diameter. End effects were small at this location. The measurements were taken at the intermediate axial point between two screw locations.

Ambient field measurements are shown in Table 2. Calculated and measured values are shown in Figure 6.

| B Component of Ambient Field | Average Value [milliGauss] |
|---|---|
| $B_{axial}$ | 170 |
| $B_{horizontal}$ | 11 |
| $B_{vertical}$ | 437 |

Table 2.   Ambient Field Conditions for the Measurements

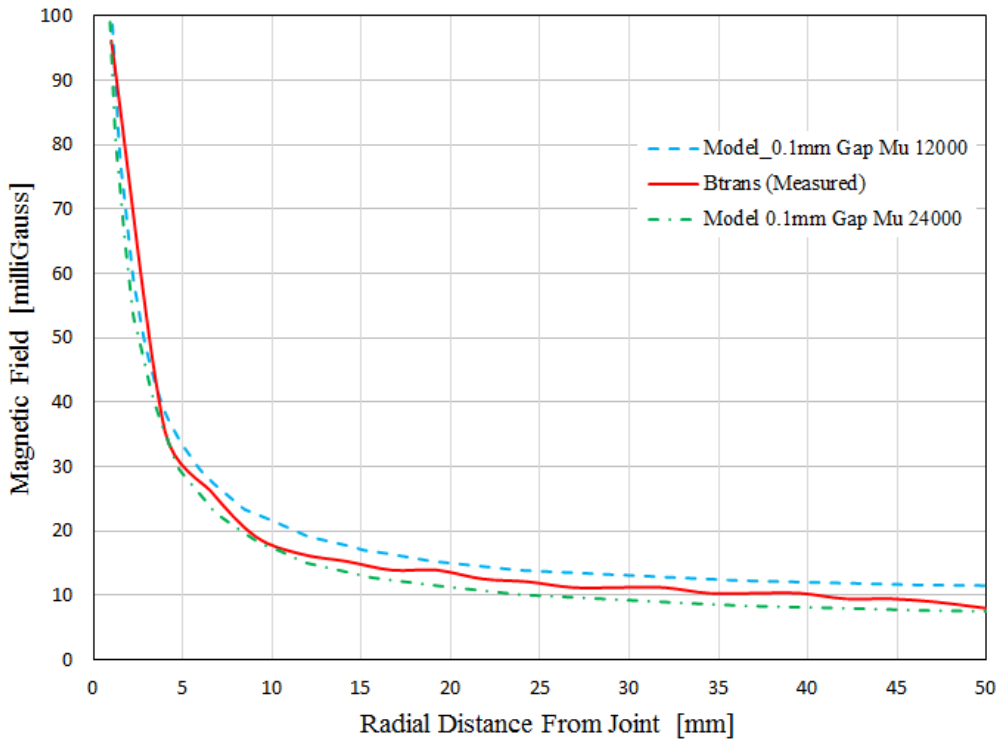

Figure 6.   Calculated and Measured Values of transverse Field for the Case of Vertical 437 mGauss Ambient Field and 0.1 mm Gap

As can be seen in Figure 6, a model gap size of 0.1 mm agrees well with the measured field profile along the radius of the cylinder. Varying the permeability of the shield material in the model does not have a large effect on the distance that field penetrates the joint. O.1 mm appears to be a reasonable value to adopt for gap size.

We next examine the case for ambient field that is rotated 90 degrees about the cylindrical axis. Calculated results are shown in Figures 7 and 8. Measurements are shown in Figure 9.

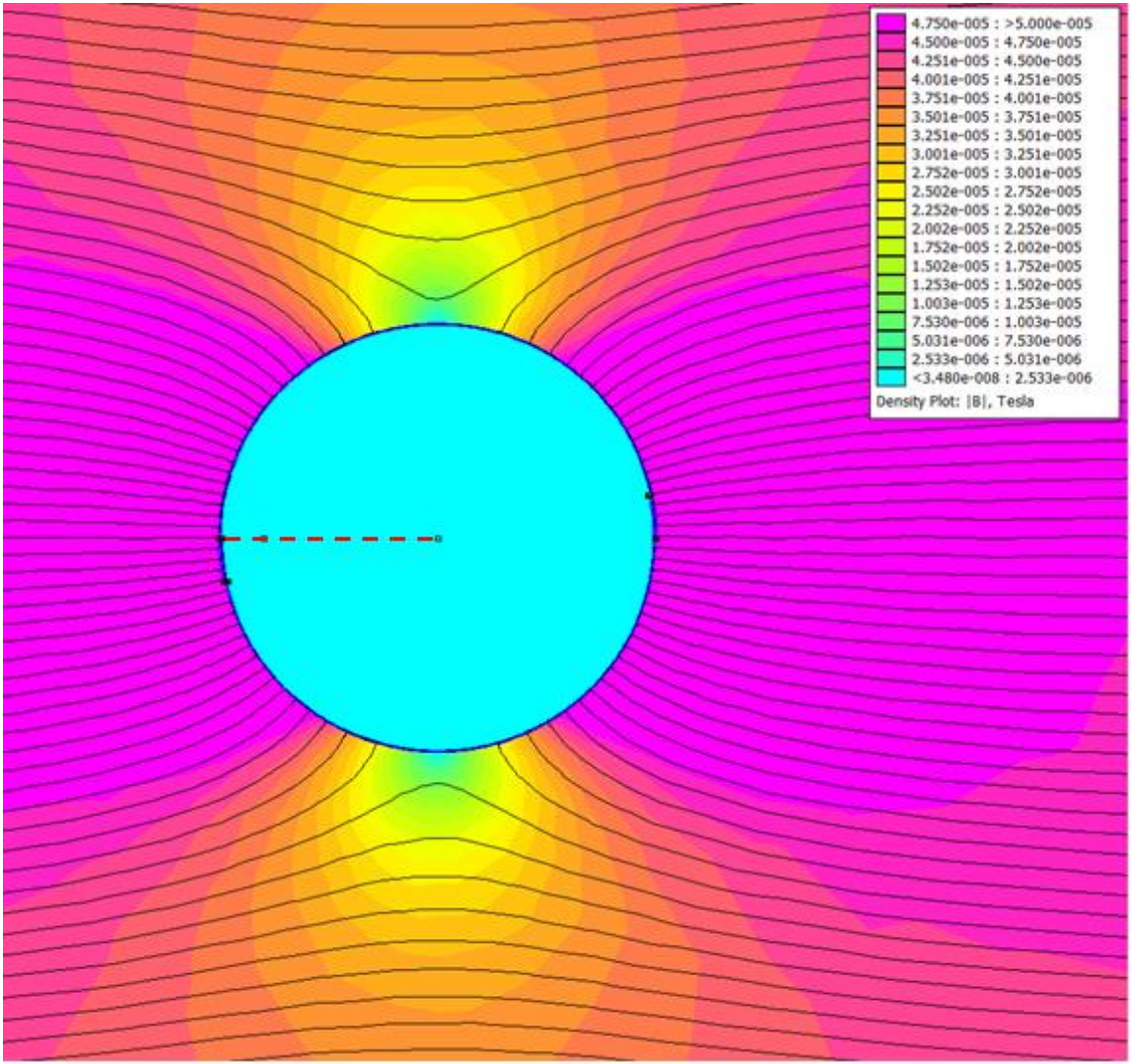

Figure 7.  The Case of Immersion in a 437 milliGauss Horizontal Ambient Field
(Calculation for 0.1 mm Gap, $\mu_r$ = 24000 shown)

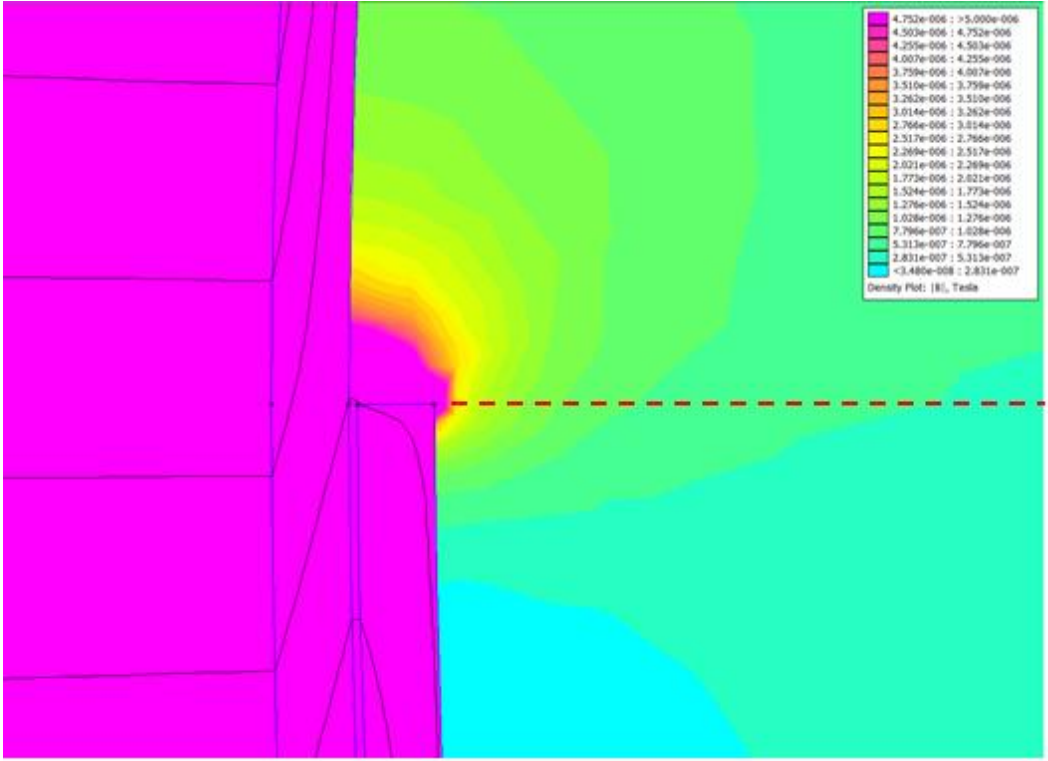

Figure 8. Enlarged View of the Joint Area From Figure 7

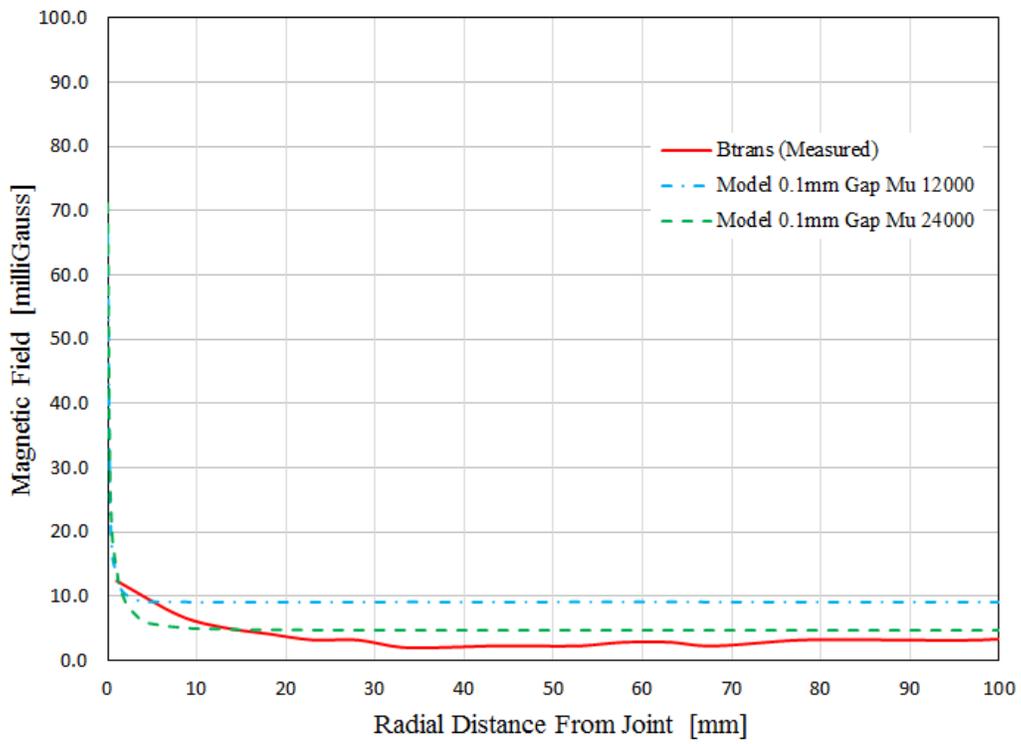

Figure 9. Calculated and Measured Values for the Case of Horizontal 437 mGauss Ambient Field and 0.1 mm Gap

The leakage field is so low in Figure 9, approximating the field for a continuous cylindrical profile for points far away from the joint, that it allows us to attempt an evaluation of the low field steady state relative permeability of the shield material (Cryoperm10, annealed for 4K operation) at room temperature. The best fit of data to the model is achieved with a relative permeability of approximately 24,000. This technique for evaluating permeability by measuring attenuation has large uncertainty. Since attenuation of the cylindrical shield in the transverse direction is approximately proportional to the permeability, a measurement error of 2 mGauss for the approximately 4 mGauss attenuated field near the center of the cylinder will result in up to a factor of two error in the calculated permeability. ±2 mGauss is the approximate uncertainty for these measurements.

**Applications**

For a Fermilab ILC cavity [2], the radial distance from the magnetic shield inner diameter and the cell outer diameter near the equator is approximately 16 mm. It can be seen in Figure 6 that the field leaking through the joint gap has decayed significantly in a distance of 16 mm, and does not present a problem for the field magnitude and direction shown. It is best to orient the location of joints in cylindrical shields so that a field distribution similar to Figure 7 is obtained. In practice, this can be difficult due to the complexity of mechanical attachments to the cavity helium vessel within the shield.

**Conclusions**

1. It is reasonable to use an average gap width of 0.1mm when modeling overlap joints in cylindrical shields.
2. It is reasonable to use an average relative permeability of 24,000 ± 12,000 for Cryoperm10 at room temperature for the purpose of modeling.